\documentclass[prl, twocolumn, superscriptaddress, floatfix, amsmath,amssymb, aps]{revtex4-2}
\usepackage{graphicx}
\usepackage{dcolumn}
\usepackage{bm}
\usepackage{hyperref}
\usepackage{siunitx}
\DeclareSIUnit\gauss{G}
\usepackage{braket}
\usepackage[normalem]{ulem}
\usepackage[dvipsnames]{xcolor}
\usepackage{tabularx}
\usepackage{xparse,xcoffins}
\ExplSyntaxOn
\NewCoffin\imagecoffin
\NewCoffin\labelcoffin
\keys_define:nn { figs/label }
 {
  label   .tl_set:N = \l_label_tl,
  labelbox .bool_set:N = \l_label_box_bool,
  labelbox .default:n = true,
  fontsize .tl_set:N = \l_label_size_tl,
  fontsize .initial:n = \footnotesize,
  pos .choice:,
  pos/nw .code:n = \tl_set:Nn \l_label_pos_tl { left,up },
  pos/ne .code:n = \tl_set:Nn \l_label_pos_tl { right,up },
  pos/sw .code:n = \tl_set:Nn \l_label_pos_tl { left,down },
  pos/se .code:n = \tl_set:Nn \l_label_pos_tl { right,down },
  pos/n .code:n = \tl_set:Nn \l_label_pos_tl { hc,up },
  pos/w .code:n = \tl_set:Nn \l_label_pos_tl { left,vc },
  pos/s .code:n = \tl_set:Nn \l_label_pos_tl { hc,down },
  pos/e .code:n = \tl_set:Nn \l_label_pos_tl { right,vc },
  pos .initial:n = nw,
  unknown .code:n   = \clist_put_right:Nx \l_label_clist
                       { \l_keys_key_tl = \exp_not:n { #1 } }
 }
\clist_new:N \l_label_clist
\box_new:N \l_label_box
\box_new:N \l_label_image_box

\NewDocumentCommand{\xincludegraphics}{O{}m}
 {
  \group_begin:
  \tl_clear:N \l_label_tl
  \clist_clear:N \l_label_clist
  \keys_set:nn { figs/label } { #1 }
  \tl_if_empty:NTF \l_label_tl
   {
    \includegraphics:Vn \l_label_clist { #2 }
   }
   {
    \SetHorizontalCoffin\imagecoffin
     {
      \includegraphics:Vn \l_label_clist { #2 }
     }
    \SetHorizontalCoffin\labelcoffin
     {
      \raisebox{\depth}
       {
        \bool_if:NTF \l_label_box_bool
         { \fcolorbox{white}{white}{\l_label_size_tl\l_label_tl} }
         { \l_label_size_tl\l_label_tl }
       }
     }
    \SetVerticalPole\imagecoffin{left}{3pt+\CoffinWidth\labelcoffin/2}
    \SetVerticalPole\imagecoffin{right}{\Width-3pt-\CoffinWidth\labelcoffin/2}
    \SetHorizontalPole\imagecoffin{up}{\Height-3pt-\CoffinHeight\labelcoffin/2}
    \SetHorizontalPole\imagecoffin{down}{3pt+\CoffinHeight\labelcoffin/2}
    \use:x{\JoinCoffins\imagecoffin[\l_label_pos_tl]\labelcoffin[vc,hc]} 
    \TypesetCoffin\imagecoffin
   }
   \group_end:
 }
\NewDocumentCommand{\setlabel}{m}
 {
  \keys_set:nn { figs/label } { #1 }
 }
\cs_new_protected:Nn \includegraphics:nn
 {
  \includegraphics[#1]{#2}
 }
\cs_generate_variant:Nn \includegraphics:nn { V }
\ExplSyntaxOff
\DeclareMathOperator{\arctantwo}{arctan2}

\usepackage{braket}

\begin{document}

\title{Interactions of electrons and Rydberg excitons in two-dimensional semiconductors}

\author{Arthur Christianen}
\affiliation{Institute for Quantum Electronics, ETH Zürich, Zürich, Switzerland}
\affiliation{Institute for Theoretical Physics, ETH Zürich, Zürich, Switzerland}
\author{Anna M. Seiler}
\affiliation{Institute for Quantum Electronics, ETH Zürich, Zürich, Switzerland}
\author{Alperen T\"uğen}
\affiliation{Institute for Quantum Electronics, ETH Zürich, Zürich, Switzerland}
\author{Ata{\c{c}} {\.I}mamo{\u{g}}lu}
\affiliation{Institute for Quantum Electronics, ETH Zürich, Zürich, Switzerland}
\date{September 23, 2026}

\begin{abstract}Rydberg excitons in two-dimensional semiconductors provide sensitive and non-destructive probes of physics in proximal sample layers that host correlated electronic states. In particular, electron or hole doping of the sample layer is heralded by a strong frequency shift and loss of transition strength of 2s excitons in the sensor layer. 
In analogy to the attractive polaron observed near the 1s exciton, this redshifted resonance, which we term Rydberg attractive polaron (RAP), has been attributed to the formation of a bound state of a 2s exciton and a remote electron. Through a theoretical analysis of exciton-electron scattering, we show that the nature of the RAP and its underlying trion can only be explained by electron-mediated hybridization of 2s, 2p and interlayer excitons, endowing the RAP a different structure than the 1s attractive polaron. We anticipate that this new understanding will ensure a more accurate assessment of the signatures of correlated electrons in two dimensional materials. 
\end{abstract}

\maketitle

Exciton spectroscopy in two-dimensional (2D) semiconductors, particularly transition metal dichalcogenides (TMDs), has emerged as a powerful tool for characterizing electronic phases of matter \cite{xu:2020, smolenski:2021,zeng:2023,cai:2023,kiper:2025}. The use of Rydberg excitons, excited excitonic states with large spatial extent, offers a promising avenue for remote sensing of materials placed in proximity to the host 2D layer \cite{xu:2020,popert:2022,zhang:2022,he:2024,mhenni:2024,tuugen:2025,hu:2025,xie:2025}. In particular, this technique has been used to detect Mott-Wigner states in semiconductor Moiré materials \cite{xu:2020} and fractional quantum Hall states in graphene \cite{popert:2022}.

Upon doping of carriers into a TMD system, a red shifted resonance appears next to the neutral exciton peak in the optical reflection or photoluminescence spectra \cite{mak:2013}. It is well established that this resonance originates from a bound state of an electron (or hole) and an exciton and is termed a trion~\cite{kylanpaa:2015,vanderdonck:2017,courtade:2017,fey:2020,klein:2022,kezerashvili:2024,perea-causin:2024,christianen:2025}. In a more complete many-body description \cite{sidler:2017,efimkin:2017,massignan:2025}, the exciton is dressed by particle-hole excitations from the bath formed by the carriers, giving rise to attractive and repulsive polarons. The attractive polaron (AP) corresponds to an exciton dressed by collective trion excitations, and at low carrier densities its energy approaches the trion energy. The repulsive polaron connects to the neutral exciton peak and blueshifts as a function of density.

For 2s excitons, a peak seemingly similar to the 1s attractive polaron appears in the spectrum upon electron or hole doping of either the same monolayer \cite{wagner:2020,sell:2022,liu:2021,biswas:2023} or a nearby TMD layer \cite{xu:2020,zhang:2022,mhenni:2025}. We refer to this feature as Rydberg attractive polaron (RAP). 
As in the 1s exciton case, this state can be understood as the exciton dressed by collective trion excitations.

\begin{figure}[!b]
    \centering
    \xincludegraphics[width=0.8\linewidth, label=a)]
    {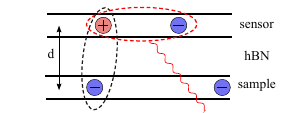}
    \xincludegraphics[width=0.8\linewidth, label=b)]{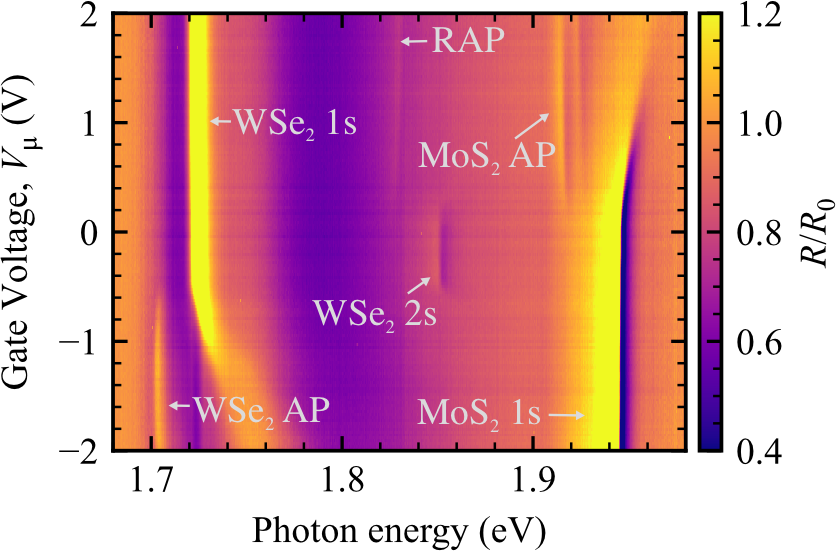}
    \caption{a) Sketch of the interlayer Rydberg trion state, formed by a carrier in the sample layer, and a Rydberg exciton in the sensor layer. b) Representative normalized reflectance  \(R/R_{0}\) spectrum for this scenario, as a function of photon energy and gate voltage \(V_{\mu}\). In this device (Device 3, see SM) a sensor WSe$_2$ layer is separated by 1.7--2.3 nm thick hBN from a sample MoS$_2$ layer. At positive (negative) gate voltages, electrons (holes) are introduced in the MoS$_2$ (WSe$_2$) layer. The 1s exciton and AP features are shown for both layers, and the 2s WSe$_2$ peak and the RAP appear in the middle.}
    \label{fig:intro_fig}
\end{figure}

Here we investigate the nature of the Rydberg trion states underlying this RAP feature. In  Fig.~\ref{fig:intro_fig}a) such a Rydberg trion is drawn for the interlayer case.
In Fig.~\ref{fig:intro_fig}b), we show a representative reflectance spectrum from an unaligned WSe$_2$/MoS$_2$ bilayer, displaying the WSe$_2$ 1s, 1s AP and 2s resonances, the MoS$_2$ 1s and 1s AP resonances, as well as the RAP.

We theoretically show that the Rydberg trions underlying the RAPs have a qualitatively different structure from 1s trion states: the Rydberg trions cannot be understood as originating simply  from the binding of a charge carrier to the 2s exciton. 
In fact, the 2s exciton has repulsive interactions with charge carriers. 
Instead, the Rydberg trion much more closely resembles a 2p exciton or an interlayer exciton (IX) state loosely bound to an additional electron. Its oscillator strength follows from the finite overlap with the 2s exciton state, due to the electron-induced hybridization between the excitonic states.
The energy splitting between the 2s exciton and the RAP peak is therefore largely determined by the 2s-2p or 2s-IX splitting. Hence, this feature gives direct spectroscopic access to these otherwise dark excitonic states. Recent work~\cite{kim:2025} highlighted the importance of 2s and 2p hybridization for understanding exciton spectra in the case of a generalized Wigner crystal in a proximal sample layer~\cite{xu:2020}, by modeling electrons as static charges. 
Our work generalizes this result to show that dynamic electron-exciton correlations in homogeneous or Moiré systems with mobile electrons also leads to hybridization of the exciton states, and that this underlies the RAP resonance which occurs ubiquitously in both monolayer and bilayer TMDs.

We start our analysis by explaining why the 2s exciton has repulsive interactions with electrons and holes. Then we will compute effective exciton-electron potentials for both the monolayer and bilayer cases. After explaining the origin of the experimental signature and the nature of the probed states, we conclude and give an outlook.

\textit{Exciton polarizability.-} We first consider a monolayer TMD encapsulated in hexagonal boron nitride (hBN) and we use atomic units. We assume parabolic valence and conduction bands and we neglect quantum geometry effects and electron-hole exchange interactions. 
We model the Coulomb interactions in the TMD
using a Rytova-Keldysh potential \cite{rytova:1967,keldysh:1979,goryca:2019,fey:2020}
\begin{equation}
V_{RK}(r)= \frac{\pi}{2 r_0}\left[H_0\left(\frac{\kappa r}{r_0} \right)-Y_0\left( \frac{\kappa r}{r_0}\right) \right].
\end{equation}
 Here, $H_0$ and $Y_0$ are the zeroth-order Struve function and Bessel function of the second kind. The values of $\kappa\approx 4.4$ and $r_0$ for TMDs have been fitted from exciton spectra in Ref.~\cite{goryca:2019}. Asymptotically, the potential scales as the bare Coulomb interaction $\frac{1}{\kappa r}$.

Since intralayer excitons do not have a permanent dipole moment, the leading-order exciton-electron interaction depends on the exciton polarizability $\alpha$, and is asymptotically given by \cite{fey:2020}
\begin{equation}
    V_{\alpha}(R) = -\frac{\alpha}{2} \left(\frac{dV_{RK}}{dR}\right)^2 \sim R^{-4} \  \mathrm{for} \ \ R \rightarrow \infty.
\end{equation}
The polarizability of an exciton in state $i$ can be computed from second-order perturbation theory:
\begin{equation}\label{eq:polarizability}
\alpha_i =2 \sum_{n\neq i}\frac{|\langle n | \hat{\bm{d}} |i \rangle|^2}{E_n-E_i},
\end{equation}
where $n$ sums over all states of the exciton other than $i$ with energies $E_n$. The dipole operator $\hat{\bm{d}}$ couples states with an angular momentum difference of $\pm1$. We show the resulting polarizabilities for various exciton states in Table \ref{tab:polarizabilities}, together with the exciton binding energies and radial extent. 

Notably, the 2s exciton has a negative polarizability, indicating a repulsive interaction with charge carriers. This behavior arises from the strong dipole coupling between the 2s and 2p exciton states, the latter lying lower in energy. The resulting negative contribution from the 2p states outweighs the positive contributions from higher-lying p states, leading to an overall negative result. Moreover, the polarizabilities of the 2s and 2p excitons are one to two orders of magnitude larger than that of the 1s state due to their greater spatial extent. Consequently, these excited excitons are more sensitive to external electric fields or the presence of charge carriers, which accounts for their effectiveness as remote sensors.

\begin{table}[t]
    \centering
    \caption{Exciton binding energies, sizes and polarizabilities for hBN-encapsulated monolayer TMDs, calculated based on material parameters from Ref.~\cite{goryca:2019} For the 2p-excitons we compute the polarizability for the 2p+ and 2p- states, 
    assuming a small splitting between these states that lifts their degeneracy. %
    The polarizability is given in atomic units.}
    \begin{tabular}{c|c c c}
    \hline
    \hline
         Exciton& $E_{\mathrm{bind}}$ (meV) & $\sqrt{\langle r^2 \rangle}$ (nm) & $\alpha$ ($10^5$ a.u.)  \\
         \hline
         MoSe2 1s& 232.0 & 1.10 & 0.71 \\
         MoSe2 2p& 79.0 & 2.97  & 29.6 \\
          MoSe2 2s& 60.6 & 4.32  & -6.87\\
        WSe2 1s& 167.0 & 1.66 & 2.06 \\
         WSe2 2p& 50.3 & 4.74  &  120.9 \\
          WSe2 2s&  38.9&  6.85  &  -51.6\\
         \hline
         \hline
    \end{tabular}
    \label{tab:polarizabilities}
\end{table}

\textit{Effective exciton-electron interaction potentials.-}
Due to the negative polarizability of the 2s state, 
 a charge carrier can not directly bind to the 2s-exciton to form a deeply bound trion state or attractive polaron state. To corroborate this further, we construct effective electron-exciton potentials and compute trion energies using the adiabatic hyperspherical approach \cite{macek:1968,greene:2017}.

\begin{figure}[!t]
    \centering
    \includegraphics[width=\linewidth]{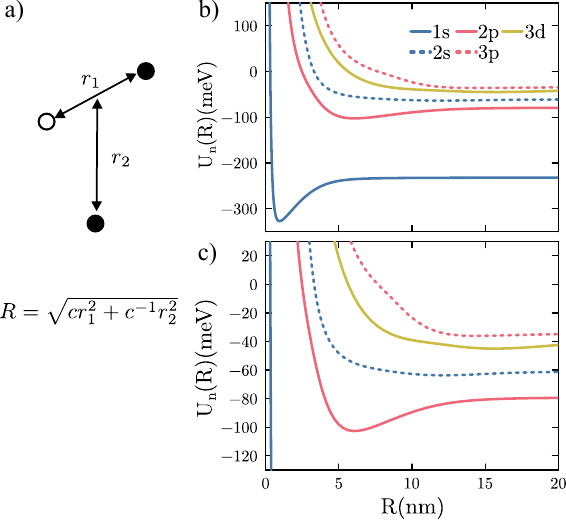}
    \caption{a) Definition of the Jacobi coordinates and the hyperradius. b)  
    Hyperspherical curves for electron-exciton scattering in MoSe$_2$ in the singlet sector, with $M=0$ and $r=0$ (see SM).
    The zero of energy is given by the energy of the free electrons and hole. 
    c) Zoom in for the higher-lying hyperspherical curves in b).}
    \label{fig:hypcurves_mono}
\end{figure}

The adiabatic hyperspherical approach is a technique for solving the three-body problem, which gives particular insight into the nature of the interactions between the collision partners \cite{macek:1968,greene:2017}. This method has been formulated in two dimensions by D'Incao and Esry \cite{dincao:2014}, and we follow their implementation. 
Central to the hyperspherical approach is the definition of the hyperradius $R=\sqrt{c r_1^2 + c^{-1} r_2^2}$
which follows from the mass-weighted combination of the two Jacobi-distances as depicted in Fig.~\ref{fig:hypcurves_mono}a), such that it is independent of the chosen set of Jacobi-coordinates. The coefficient $c=\sqrt{\frac{\mu_X}{\mu}}$, where $\mu_X=\frac{m_1 m_2}{m_1+m_2}$ and $\mu=\sqrt{\frac{m_1 m_2 m_3}{m_1+m_2+m_3}}$.   A first description of trions in TMDs using hyperspherical coordinates was carried out in Ref.~\cite{kezerashvili:2024}.

In terms of these hyperspherical coordinates, the time-independent Schr{\"o}dinger equation becomes
\begin{equation}
\left[-\frac{1}{2\mu}\frac{\partial^2}{\partial R^2}+\frac{\hat{\Lambda}^2+\frac{3}{4}}{2\mu R^2}+V(R,\Omega)\right] \Psi(R,\Omega)=E \Psi(R,\Omega),
\end{equation}
where $\Omega$ denotes the set of hyperangles and $\hat{\Lambda}$ is the grand angular momentum operator given in Ref.~\cite{dincao:2014}. In the \textit{adiabatic} hyperspherical approach one first solves the hyperangular equation for every hyperradius
\begin{equation}
\left[\frac{\hat{\Lambda}^2+\frac{3}{4}}{2\mu R^2}+V(R,\Omega)\right] \Phi_n(R,\Omega) =U_n(R) \Phi_n(R,\Omega),
\end{equation}
giving adiabatic basis functions $\Phi_n(R,\Omega)$. In terms of this basis, the total wave function can be written as
\begin{equation}
    \Psi(R,\Omega)= \sum_nF_n(R) \Phi_n(R,\Omega),
\end{equation}
yielding coupled-channels equations for the functions $F_n(R)$, see the Supplementary Material (SM) \cite{SM}, which includes the additional references \cite{wang:2011,brem:2019,glazov:2020}. The functions $U_n(R)$ can be interpreted as effective potentials for the electron-exciton scattering for the given initial scattering channel $n$. The different channels are coupled via non-adiabatic matrix elements following from the $R$-dependence of $\Phi_n(R)$. Since the total angular momentum $M$ is conserved, this scattering problem can be solved separately for every $M$.

\textit{Monolayer results.-} We first consider RAP signatures in monolayers \cite{wagner:2020,sell:2022,liu:2021,biswas:2023}. In Fig.~\ref{fig:hypcurves_mono}b) and c), the hyperspherical curves $U_n(R)$ are shown for 
M=0
exciton-electron scattering in the singlet configuration between the electron in the exciton and the free electron. The curves are colored according to the asymptotic states they connect to. The 1s-curve is well-separated from the other curves, which are shown enlarged in Fig.~\ref{fig:hypcurves_mono}c). 

The potential adiabatically connected to the 2s threshold has only a minimal attractive well \footnote{The hyperspherical potential still has an attractive contribution from the attractive centrifugal potential characteristic of two-dimensional scattering (see SM). The actual interaction potential is repulsive.}, as expected from the negative polarizability. Therefore, this adiabatic potential cannot host a bound state with a binding energy of 10 to 20 meV, as experimentally observed. 

By contrast, the lower-lying potential asymptotically corresponding to the 2p state does host a bound state. Moreover, while the 1s exciton only has an angular momentum $M=0$ singlet bound state with an electron, we find that the 2p-state also hosts higher angular momentum and spin-triplet bound states.  These trions can all be optically bright due to their hybridization with the 2s and other s states, but they are predominantly of 2p character.  Concretely, for the $M=0-2$ states the s-exciton contribution to the wave function is 20-30 \%, whereas the p-exciton weight lies around 60$\%$. The precise definition and calculation of these wave function weights is discussed in the SM. 

We compute the trion binding energies and autodetachment linewidths $\Delta E$ and tabulate them in Table \ref{tab:binding_2p_mono}. See the SM for more details on the calculation and a discussion on auto-detachment. Since the trion binding energy is small with respect to the 2p-threshold, the measured splitting between the 2s exciton and the associated trion state is dominated by the 2s-2p splitting.
For WSe$_2$, we find three trions, which are split from the 2s state by 17.5, 16.9 and 14.7 meV
, consistent with the experimentally observed 14.1 meV \cite{wagner:2020}. Based on the match with the experimental energy and the lack of direct 2s bound states, we conclude that the attractive polarons originating from these 2p trion states
are the resonances which are experimentally observed \cite{wagner:2020,sell:2022,liu:2021,biswas:2023}.

\begin{table}[t]
    \centering
    \caption{Binding energies and autodetachment widths (see SM) of Rydberg trions in monolayer TMDs formed from an electron and a 2p exciton, depending on the total angular momentum and wave-function symmetry. The binding energy is measured with respect to the 2p exciton state. We consider the negative trion in MoSe$_2$ and the positive trion in WSe$_2$.}
    \label{tab:binding_2p_mono}
    \begin{tabular}{c c | c c |c  c}
    \hline
    \hline
    \multicolumn{2}{c}{Quantum numbers} & \multicolumn{2}{c}{MoSe$_2$} & \multicolumn{2}{c}{WSe$_2$} \\
        $M$& singlet/triplet& $E$ (meV) & $\Delta E$ & $E$ (meV) & $\Delta E$ \\
    \hline
        0 & s  & 9.3 & $6.3 \times 10^{-5}$ & 6.1 & 0.0015 \\
        $\pm$ 1 & t & 8.4 & 0.15 & 5.4 & 0.11  \\
      $\pm$ 2 & s & 5.4 & 0.034 & 3.3& 0.0064  \\
        $\pm$ 3 & t &  0.70 & 0.025 & - & -  \\
    \hline
    \hline
    \end{tabular}
\end{table}

\textit{Proximity sensing.-}
We next analyze the experimentally more interesting case where the injected Rydberg exciton resides in a different layer from the charge carriers and can be used to sense correlated electronic states. 
We consider a 2s exciton in a WSe$_2$ sensor layer, separated by a variable number of hBN spacer layers from a sample MoSe$_2$ layer that is gate-doped with electrons. We assume that there is no tunneling between the WSe$_2$ and MoSe$_2$ layers and no Moiré potential. We model the electron-hole interactions by accounting for the mutual screening between the TMD layers as in Ref.~\cite{wagner:2025}. We analyze the dependence of the exciton and trion energies on the interlayer spacing $d$, defined as the distance between the centers of the TMD layers \footnote{This distance $d$ is not included in the hyperradius, which describes the motion in the plane, but it is included in the interaction potentials.}.

\begin{figure}[!t]
    \centering
    \includegraphics[width=\columnwidth]{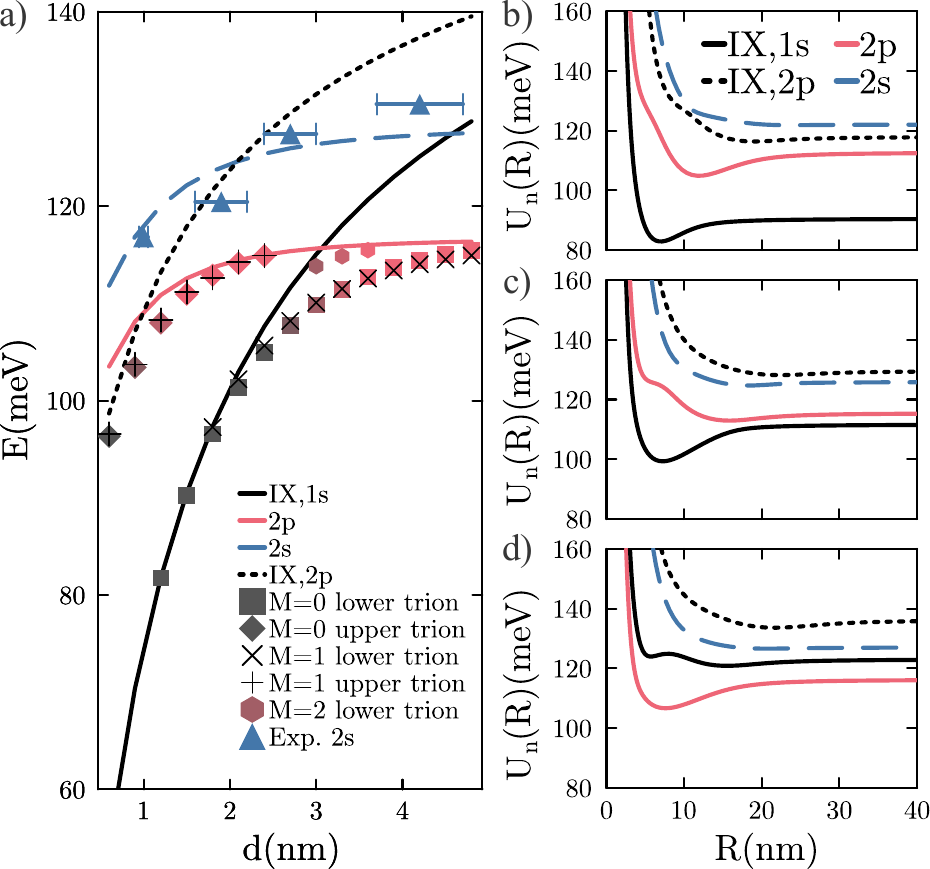}
    \caption{a) Energies of the exciton and trion states for the WSe$_2$-hBN-MoSe$_2$ heterostructure, as a function of the TMD interlayer separation $d$. We consider the configuration with an electron-hole pair in WSe$_2$ and an electron in MoSe$_2$.  The energy is measured with respect to the 1s-exciton state in the WSe$_2$ layer. b-d) Hyperspherical curves for the same setting as a) corresponding to the interlayer separations of b) 1.5nm, c) 2.7 nm and d) 3.9 nm.}
    \label{fig:Interlayer_trion_energies}
\end{figure}

Fig.~\ref{fig:Interlayer_trion_energies}a) shows the energies of the exciton (lines) and trion (markers) states with respect to the 1s exciton energy in WSe$_2$, together with the experimentally measured 1s-2s splittings from four different devices (see SM). The good agreement between the experimental and theoretical 2s energies shows that the interlayer dielectric screening is adequately accounted for. 

Interestingly, two trion branches emerge, which we will refer to as the upper and lower branches. Their nature becomes clear from the adiabatic hyperspherical curves shown for various interlayer distances in Figs.~\ref{fig:Interlayer_trion_energies}b-d). For small interlayer separation $d<2$nm both of the trion branches appear, but the lower trion is only weakly bound. In this regime, the interlayer exciton state and the 2p-exciton state are energetically well separated and the interlayer exciton interacts only weakly with the electron in the WSe$_2$ layer, as visualized with the black adiabatic potential in Fig.~\ref{fig:Interlayer_trion_energies}b). The red curve in Fig.~\ref{fig:Interlayer_trion_energies}b), asymptotically corresponding to the intralayer 2p exciton, profits from hybridization between the 2p, 2s exciton and interlayer 2p exciton states. This leads to the binding of the upper trion state of a few meV with respect to the 2p threshold.

\begin{figure}
    \centering
    \xincludegraphics[width=0.95\linewidth,label=a)]{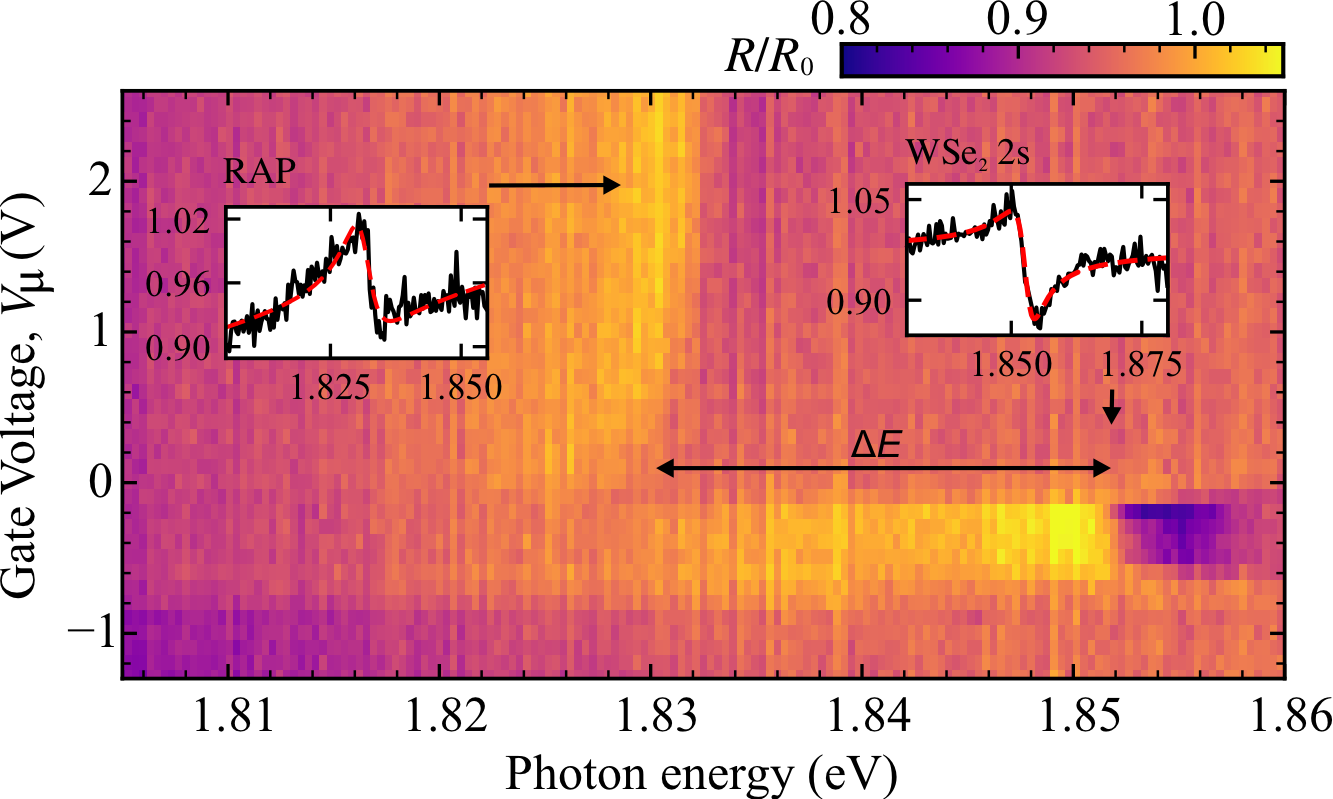}
    \vspace{0.1em}
    \xincludegraphics[width=0.95\linewidth,label=b)]{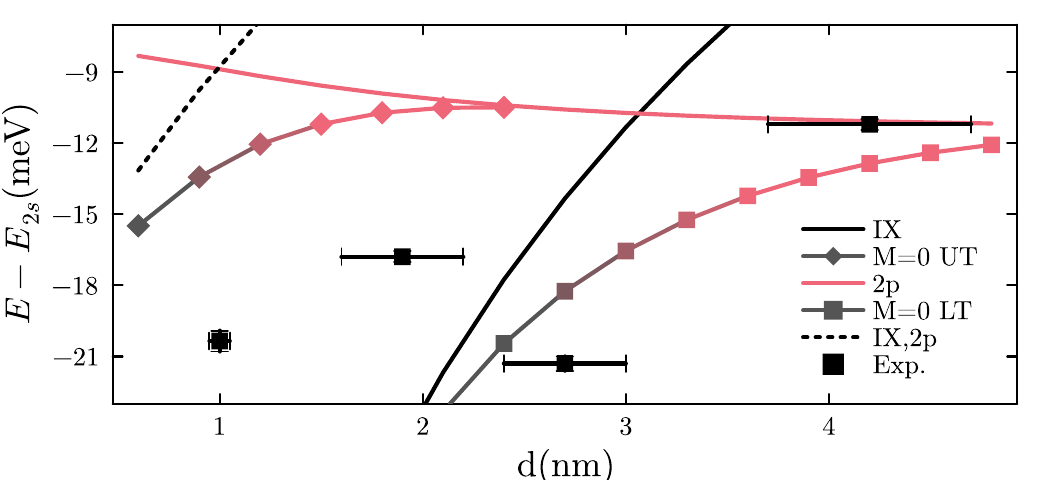}
    \caption{a) Map of the normalized reflectance \(R/R_{0}\) as a function of photon energy and gate voltage \(V_{\mu}\) for Device 3 (WSe$_2$/MoS$_2$, 1.7--2.3\,nm spacer). Insets show line cuts (black) together with fits (red, dashed) of the 2s exciton and RAP resonances. b) Experimental measurement of the RAP resonance compared to the 2s-energy, in comparison with the theoretical interlayer trion calculation (only showing the $M=0$ states).}
    \label{fig:experiment}
\end{figure}

At intermediate separations ($2\leq d \leq 3$ nm), the upper trion disappears while the lower trion binding energy increases. This enhanced binding originates from the admixing of the interlayer and the WSe$_2$ 2p exciton states, which are close in energy in this regime, leading to the deep black adiabatic potential in Fig.~\ref{fig:Interlayer_trion_energies}c). For larger distances $d>3$ nm this mixing is gradually suppressed as the interlayer exciton binding energy decreases and its splitting from the 2p state grows, causing the trion energy to approach the 2p threshold.  
The change in nature of the trions is indicated by the gradual change in color of the symbols (except for the $M=1$ excitons, for which the symbols are shown in black for visibility).

As in the monolayer case, trions with higher angular momentum also occur. The $M=1$ trions are close in energy to the $M=0$ trions throughout the studied parameter range. An $M=2$ trion only appears near the crossing point of the interlayer exciton and 2p exciton states, where trion binding is the most favorable.

\textit{Experimental results.-}
We experimentally studied several WSe$_2$–MoS$_2$ and WSe$_2$–MoSe$_2$ bilayer samples with different hBN spacer thicknesses (see SM for details).
Figure~\ref{fig:experiment}a) shows a representative reflectance spectrum for a WSe$_2$–MoS$_2$ bilayer, focusing on the 2s exciton and RAP resonances (the same data are shown in Fig.~\ref{fig:intro_fig}(b) over a broader energy range).
The insets display fits to the WSe$_2$ exciton and RAP resonances at selected carrier densities, from which the energy splittings were extracted (see SM).

We find that the splitting between the RAP and the 2s exciton generally decreases with spacer thickness, except for an abrupt increase between the second and third experimental data points. This is consistent with the theoretically predicted two trion branches, within each of which the binding energy decreases with interlayer distance. Alternatively, the jump could originate from the change from MoSe$_2$ in devices 1 and 2 to MoS$_2$ in devices 3 and 4. This appears unlikely, as our calculations predict only a few-percent difference in the 2s-RAP splitting between the two materials. However, conclusively establishing the presence of two trion branches experimentally will require measurements on additional samples.

Quantitatively, clear discrepancies between experiment and theory appear at both small and large $d$ (see also SM). One important ingredient missing from our theoretical description is the Berry curvature, which produces substantial splittings of the 2p states \cite{srivastava:2015}. Consistent with this interpretation, the best agreement with theory is found at $d=2.7$ nm, where the trion has the strongest interlayer-exciton character and the Berry-curvature-induced 2p splitting is therefore expected to play a smaller role.

\textit{Discussion and conclusion.-}
We have studied the experimentally observed signatures of Rydberg attractive polarons near the 2s state in monolayer and bilayer TMDs. In contrast to previous interpretations, we find that these features predominantly originate from trions with 2p- or interlayer-exciton character. They acquire optical brightness through hybridization with the 2s state, thereby enabling optical access to otherwise dark excitonic states. In WSe$_2$-MoSe$_2$ bilayers, we find not one but two trion branches as the hBN thickness is varied, both with mixed 2p- and interlayer-exciton character.

A promising method to probe the predicted structure of the Rydberg trion is angle-resolved photoemission spectroscopy (ARPES) \cite{wu:2025,meneghini:2026}. By ejecting one of the trion's electrons and resolving the resulting photoemission channels, ARPES could reveal the relative 2p- and interlayer-exciton contributions to the trion wave function.

An open challenge is to understand the unusual density dependence of the RAP peak shown in Fig.~\ref{fig:experiment}a), in particular its sharp blueshift at low densities. This behavior differs markedly from that of the 1s AP and reflects the distinct nature of the RAP. Band-gap renormalization and the screening-induced reduction of exciton binding are expected to play an important role in the RAP line shift \cite{wolf:2025}. However, a full description of the density-dependent RAP properties will require a theory that explicitly incorporates both many-body dressing and the internal structure of the Rydberg trion, providing an interesting direction for future work.

\textit{Acknowledgements-} The authors acknowledge useful discussions with Richard Schmidt, Rafał Ołdziejewski, Haydn Adlong, Martin Kroner, Amine Ben Mhenni and Alexey Chernikov. A. C. and A.M.S. were supported by an ETH Fellowship.

\bibliography{references}

\clearpage
\onecolumngrid


\begin{center}
{\Large\bfseries Supplemental Material}
\end{center}

\vspace{1em}

\setcounter{section}{0}
\setcounter{figure}{0}
\setcounter{table}{0}
\setcounter{equation}{0}

\renewcommand{\thesection}{S\arabic{section}}
\renewcommand{\thefigure}{S\arabic{figure}}
\renewcommand{\thetable}{S\arabic{table}}
\renewcommand{\theequation}{S\arabic{equation}}

\section{Overview of adiabatic hyperspherical approach}

\subsection{Formalism and implementation}
As presented in the main text, the goal of the adiabatic hyperspherical approach is to solve the hyperradial Schr{\"o}dinger equation
\begin{equation}
\left[-\frac{1}{2\mu}\frac{\partial^2}{\partial R^2}+\frac{\hat{\Lambda}^2+\frac{3}{4}}{2\mu R^2}+V(R,\Omega)\right] \Psi(R,\Omega)=E \Psi(R,\Omega),
\end{equation}
by first constructing the hyperangular eigenstates $\Phi_n(R)$ at every hyperradius
\begin{equation}
\left[\frac{\hat{\Lambda}^2+\frac{3}{4}}{2\mu R^2}+V(R,\Omega)\right] \Phi_n(R,\Omega) =U_n(R) \Phi_n(R,\Omega),
\end{equation}
and then propagating the Schr{\"o}dinger equation outward using the coupled-channels equations.

We follow the prescription of D'Incao and Esry \cite{dincao:2014}.
The grand angular momentum operator $\hat{\Lambda}$ is defined in terms of the hyperangles $\bm{\Omega}=(\theta,\phi,\gamma)$ as 
\begin{equation}
    \hat{\Lambda}^2(\bm{\Omega})=- 4 \left( \frac{1}{\sin \theta} \frac{\partial}{\partial \theta  }  \sin \theta \frac{\partial}{\partial \theta  } + \frac{1}{\sin^2 \theta} \frac{\partial^2}{\partial \phi^2}\right)- \frac{1}{\sin^2 \theta} \left(\frac{\partial^2}{\partial \gamma^2}-4 \cos \theta \frac{\partial^2}{\partial \gamma \partial \phi} \right),
\end{equation}
where $\gamma$ is the Euler angle specifying the overall orientation of the complex, and $\phi$ and $\theta$ can be defined in terms of the interparticle distances given by
\begin{align}
r_{12}&=\frac{1}{\sqrt{2}} d_{12} R \sqrt{1+\sin \theta \cos(\phi+\phi_{12})}, \\
r_{23}&=\frac{1}{\sqrt{2}} d_{23} R \sqrt{1+\sin \theta \cos(\phi+\phi_{23})}, \\
r_{31}&=\frac{1}{\sqrt{2}} d_{31} R \sqrt{1+\sin \theta \cos(\phi+\phi_{31})}.
\end{align}
The quantities $d_{ij}$, and $\phi_{ij}$ are given by
\begin{align}
    \mu d_{ij}^2&=\frac{m_k (m_i +m_j)}{m_1+m_2+m_3}, \\
    \phi_{12} &=2 \tan^{-1} (m_3/\mu), \\
    \phi_{23} &=0, \\
    \phi_{31} &=-2 \tan^{-1} (m_2/\mu). 
\end{align}
For a detailed discussion on the boundary conditions and the symmetries that these coordinates should obey, see Ref.~\cite{dincao:2014}.

To solve the hyperangular equations we represent the wave function in terms of a grid of basis splines for both coordinates
\begin{equation}
    \Phi_n(R,\Omega) = \sum_{i,j} \psi_{ij} B_{i}(R,\theta) B_{j}'(R,\phi),
\end{equation}
where $B_i$ and $B_j'$ form a basis of splines over the range of $\theta$ and $\phi$. The boundary conditions are built in by constructing the basis splines with the appropriate constraints. For an efficient implementation, especially for large hyperradii, it is important to adaptively adjust the grid of splines such that a high resolution in the region of the interaction potential is maintained. Indeed, for large distances, the range of hyperangles over which bound state wave functions are extended asymptotically goes to zero. We achieve convergence to the 0.01 meV level for the shown adiabatic hyperspherical curves with 30 splines in the $\theta$ grid and 60-72 splines in the $\phi$ grid (depending on the monolayer or bilayer case).

The total scattering wave function can now be expressed in terms of these hyperradial functions
\begin{equation}
    \Psi(R,\Omega)= \sum_nF_n(R) \Phi_n(R,\Omega),
\end{equation}
leading to the coupled channels equations \begin{equation}\label{eq:hyperradialSE}
\left[-\frac{1}{2\mu}\frac{\partial^2}{\partial R^2}+U_n(R) -E\right] F_n(R) \\ -\frac{1}{2\mu} \left[ \sum_{n'}2 P_{n,n'}(R) \frac{\partial}{\partial R}+Q_{n,n'}(R)\right] F_{n'}(R) =0,
\end{equation}
where
\begin{align}
    P_{n,n'}(R)&=\int d\Omega \ \Phi_n(R,\Omega)^{\ast} \frac{\partial}{\partial R} \Phi_{n'}(R,\Omega), \\
    Q_{n,n'}(R)&=\int d\Omega \ \Phi_n(R,\Omega)^{\ast} \frac{\partial^2}{\partial R^2} \Phi_{n'}(R,\Omega)=-\int d\Omega \ \frac{\partial\Phi_n(R,\Omega)^{\ast}}{\partial R} \frac{\partial \Phi_{n'}(R,\Omega)}{\partial R}. \label{eq:Qdef}
\end{align}
These $P$ and $Q$ matrices are calculated numerically by finite difference in $R$.

\subsection{Symmetries}

For a practical implementation of the adiabatic hyperspherical approach it is important to consider the symmetries of the problem, see Ref.~\cite{dincao:2014} for a careful discussion. It is most convenient to divide the problem into symmetry sectors labeled by the magnitude of the total angular momentum $M$, the exchange symmetry of the identical particles ($\pm 1$ depending on whether they form a spin singlet or triplet), and the reflection symmetry $r=\pm1$. 

Interestingly, aside from the trion resonances shown in Table 2 in the main text, there is one more bound trion from the 2p state. This has quantum numbers $M=0$, $r=-1$ and a spin triplet configuration. This state has binding energy 1.9 meV for MoSe$_2$ and $1.05$ meV for WSe$_2$, and is a true bound state, since in this symmetry sector there is no coupling to the s-states of the exciton. Therefore, these states are optically dark from this symmetry consideration, and not shown in the main text.

\begin{figure}
    \centering
    \includegraphics[width=0.3\linewidth]{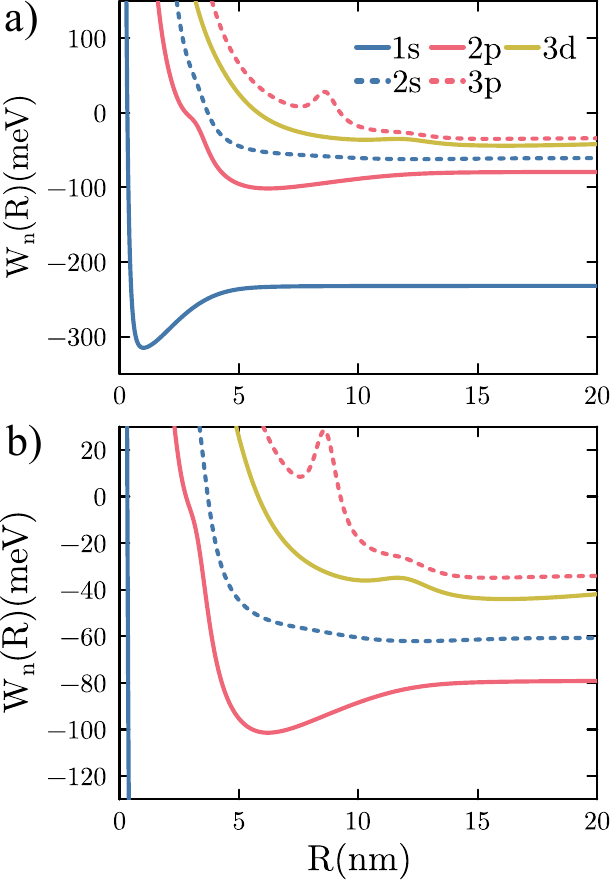}
    \caption{ a) Adiabatic hyperspherical curves $W_n(R)$ for exciton-electron scattering in MoSe$_2$, for $M=0$ and $r=0$. In contrast to the results in Fig.~2b) and c) in the main text, here the diagonal corrections $-\frac{Q_{n,n}(R)}{2\mu}$ are included. In b) a zoom-in is shown of the higher-lying curves. }
    \label{fig:Wcurves}
\end{figure}

\subsection{Asymptotic behavior of the adiabatic hyperspherical curves}

The asymptotic behavior of the adiabatic hyperspherical curves is determined both by the interaction potential and the centrifugal contributions. The correct asymptotic behavior of the adiabatic potentials is not fully captured in $U_n(R)$: also the diagonal non-adiabatic contributions from the $Q$-matrix are important. While the diagonal component $P_{n,n}(R)=0$, the value of $Q_{n,n}(R)$ is typically finite. This diagonal component of Q is often absorbed in the adiabatic potentials, defining:
\begin{equation}
    W_{n}(R)=U_n(R)-\frac{Q_{n,n}(R)}{2\mu}.
\end{equation}
From Eq.~\eqref{eq:Qdef} we see that $Q_{n,n}(R)$ is strictly negative, and that it therefore has a positive contribution to $W_{n}(R)$. These $W_{n}(R)$ capture the correct asymptotic dependence of the adiabatic potentials \cite{dincao:2014}. The $W_n(R)$ curves are presented in Fig.~\ref{fig:Wcurves} for the same parameters as in Fig.~1 in the main text, where the $U_n(R)$ curves are shown. While in the long range the difference between the $W_n(R)$ and the $U_n(R)$ curves is difficult to distinguish by eye, the $W_n(R)$ curves display clearly distinguishable additional peaks in the short range. These follow from peaks in $Q_{n,n}(R)$ caused by avoided crossings between the hyperspherical potentials: if the nature of the channel wave functions rapidly changes around a certain hyperradius, the non-adiabatic coefficients will take a large value.

For large $R$, the $W_n(R)$ curves are given by \cite{dincao:2014}
\begin{equation} \label{eq:asymp}
    W_n(R)=E_{X,n} + \frac{(M-M_{X,n})^2-\frac{1}{4}}{2 \mu R^2} + V_n(R).
\end{equation}
Here $E_{X,n}$ and $M_{X,n}$ are respectively the excitonic energy and angular momentum to which the asymptotic curve connects, and $M$ is the total angular momentum. The function $V_n(R)$ is the asymptotic interaction. The second term in Eq.~\eqref{eq:asymp} is the centrifugal potential. In two-dimensions the centrifugal potential is attractive for s-wave scattering and given by $-(8 \mu R^2)^{-1}$. This attractive centrifugal potential makes that the hyperspherical curve corresponding to scattering between the 2s exciton and an electron is slightly attractive at long range (see Figs.~1 and \ref{fig:Wcurves}), even though the potential $V_n(R)$ is repulsive. This attractive centrifugal term does not correspond to a physical attraction, since it is also present in the non-interacting problem.

As mentioned in the main text, the most important contribution to $V_n(R)$ is the charge-induced dipole interaction, asymptotically scaling as $R^{-4}$. However, formally, the most long-ranged contribution is the charge-quadrupole interaction which scales as $R^{-3}$. In two dimensions, the exciton has a quadrupole moment if the electron and hole masses are different, scaling as $\frac{m_h^2-m_e^2}{(m_e+m_h)^2}$. Although the presence of this term affects the long-range adiabatic hyperspherical curves, we did not discuss it in the main text since the quadrupole moment is relatively small, and it plays a very limited role for the relevant physics.

\subsection{Calculation parameters}

The TMD parameters  that we used in our calculations were taken from Ref.~\cite{goryca:2019}. Here, the values of $r_0$, $\kappa$ and the exciton reduced mass were measured. Since the electron and hole masses could not be directly extracted, we assumed a combination of electron and hole masses that yield the measured reduced mass. We find that our calculations are not very sensitive to this choice.
Concretely, for MoSe$_2$ we used the parameters: $m_h=0.6$ $m_e=0.84$, $r_0=3.9$ nm, and for WSe$_2$ we used: $m_h=0.45$ $m_e=0.36$, $r_0=4.4$ nm. The dielectric constant for the surrounding hBN was taken to be $\kappa=4.4$.

\section{Autodetachment of exciton-electron Rydberg trions}

To obtain the trion binding energy, we solve Eq.~\eqref{eq:hyperradialSE} using the R-matrix method \cite{wang:2011}. While the 1s trion state is a true bound state, the trions originating from Rydberg states can decay into a continuum of 1s exciton plus free electron or hole states. This process is called auto-detachment, although in the exciton literature it is sometimes referred to as auto-ionization \cite{wagner:2020}. The energies of these higher-lying trion states therefore need to be extracted by solving the 1s exciton + electron scattering problem. The bound state then shows up as a Fano-Feshbach resonance. We extract the bound-state energies and autodetachment linewidths $\Delta E$ by fitting the collision cross sections at the Fano-Feshbach resonances to the Fano profile
\begin{equation}
\sigma_{\mathrm{Fano}}(E)=\sigma_{\mathrm{bg}} \frac{\left[q \Delta E/2 +(E-E_{\mathrm{res}})\right]^2}{\Delta E^2/4+(E-E_{\mathrm{res}})^2}.
\end{equation}
Here $\sigma_{\mathrm{bg}}$ is the background collision cross section and $q$ is the Fano parameter. For this formula to be valid, it is assumed that the background collision cross section does not significantly vary over the linewidth of the resonance. The fitted Fano profile parameters for the Rydberg trion states in TMD monolayers are shown in Table \ref{tab:fanofeshbach}. These calculations were performed including 12 channels in the adiabatic hyperspherical calculations. While $E_{\mathrm{res}}$ and $\Delta E$ are fully converged, deviations in $\sigma_{\mathrm{bg}}$ and $q$ can still be found for an increased number of channels. 

\begin{table}[t]
    \centering%
    \label{tab:fanofeshbach}
    \caption{Fano-Feshbach parameters (excluding radiative lifetime), computed including the 12 lowest adiabatic hyperspherical curves. The resonance energy $E_{\mathrm{res}}$ is measured with respect to the 1s exciton energy. This is the kinetic energy which is released upon autodetachment.
    }
    \begin{tabular}{c c c c | c c c c}
    \hline
    \hline
        TMD & $M$& s/t & r & $E_{\mathrm{res}}$ (meV) & $\Delta E$ (meV) & $\sigma_{\mathrm{bg}}$ (nm) & q  \\
    \hline
        MoSe$_2$& 0 & s & 1 & 143.625 &  $6.3 \times 10^{-5}$ & 0.23 & -3.5 \\
        MoSe$_2$& 1 & t & 1,-1 & 144.56 & 0.15 & 0.99 & -1.5  \\
        MoSe$_2$& 2 & s & 1,-1 & 147.54 & 0.034 & 0.755 & -1.8  \\
        MoSe$_2$& 3 & t & 1,-1 & 153.63 & 0.020 & 0.36 & -2.7  \\
        WSe$_2$& 0 & s & 1 & 110.55 & 0.0015 & 0.37 & -3.4  \\
        WSe$_2$& 1 & t & 1, -1 & 111.20 & 0.11 & 1.4 & -1.5  \\
        WSe$_2$& 2 & s & 1, -1 & 113.34 & 0.0064 & 0.94 & -2.0  \\
    \hline
    \hline
    \end{tabular}
\end{table}

We find that the auto-detachment linewidths of all the resonances are small compared to the Rydberg trion binding energy. This can be qualitatively explained by the fact that the adiabatic hyperspherical curve asymptotically connected to the 1s-state, as shown in Figs.~1 and \ref{fig:Wcurves}, is far separated from all the higher-lying adiabatic curves. This means that non-adiabatic transitions between these states are suppressed. Neglecting the off-diagonal terms in Eq.~\eqref{eq:hyperradialSE} can therefore be a good approximation \cite{kezerashvili:2024} when only describing the 1s-states. 

The linewidth of the RAP that is typically measured in bilayer systems, including our devices, is comparable to that of the 2s exciton. This is consistent with the autodetachment process playing a limited role, as we find from our theoretical calculations. In monolayer systems, the RAP linewidth has been measured to be larger: about 12 meV in WSe$_2$ \cite{wagner:2020}. While this has been attributed to autodetachment, it is unclear at this point whether this is a generic feature of the RAP in monolayer TMDs or if this depends on the sample quality, or other experimental conditions. Other broadening mechanisms could potentially also explain this large linewidth. For example,  phonon-assisted intervalley scattering, which is important for the neutral 2s exciton \cite{brem:2019}, could be enhanced for the Rydberg trion states.

\section{Character of Rydberg trion wave functions}

While the adiabatic hyperspherical approach can give intuitive insights into the effective potentials and the scattering processes, it is not straightforward to quantitatively assess to which degree the 2s and 2p states contribute to the scattering wave function. To address this aspect, we use the approach from Refs.~\cite{fey:2020,christianen:2025}, writing the 
 three-body wave function in the center of mass frame as
\begin{equation}
    \psi_T(r_1,r_2,\theta, \alpha)=\frac{u(r_1,r_2,\theta)}{\sqrt{2\pi r_1 r_2}} \exp(i M \alpha).
\end{equation}
Here, $r_1$ and $r_2$ are the electron-hole bond lengths, $\theta$ is the relative angle between the vectors $r_1$ and $r_2$, and $\alpha$ is the angle specifying the overall orientation of the complex. The full equations of motion that $u$ obeys are given in Ref.~\cite{fey:2020}.

These equations are solved by expanding the wave function in terms of associated Laguerre polynomials $L^{\alpha}_n(x)$:
\begin{align}
u(r_1,r_2,\theta)
&=
\frac{1}{\sqrt{2\pi}}
\sum_{i,j,l} \,
\phi_{|M/2+l|,i}(r_1)\phi_{|M/2-l|,j}(r_2)e^{i l\theta },
\label{eq:varwf1} \\
\phi_{l,i}(r)
&=
\left(\frac{r}{\lambda}\right)^{l+1/2}
L_i^{2l}\!\left(\frac{r}{\lambda}\right)
e^{-r/(2\lambda)}
\sqrt{\frac{\Gamma(i+1)}{\lambda\Gamma(i+2l+1)}} .
\label{eq:varwf2}
\end{align}
We find this wave function by solving the generalized eigenvalue in a finite basis of these Laguerre polynomials. The scale $\lambda$ is optimized to obtain convergence with the smallest number of basis functions. Note that $l$ takes integer values for even $M$ and half-integer values for odd $M$. 

\begin{figure}[t]
    \centering
    \includegraphics[width=0.8\linewidth]{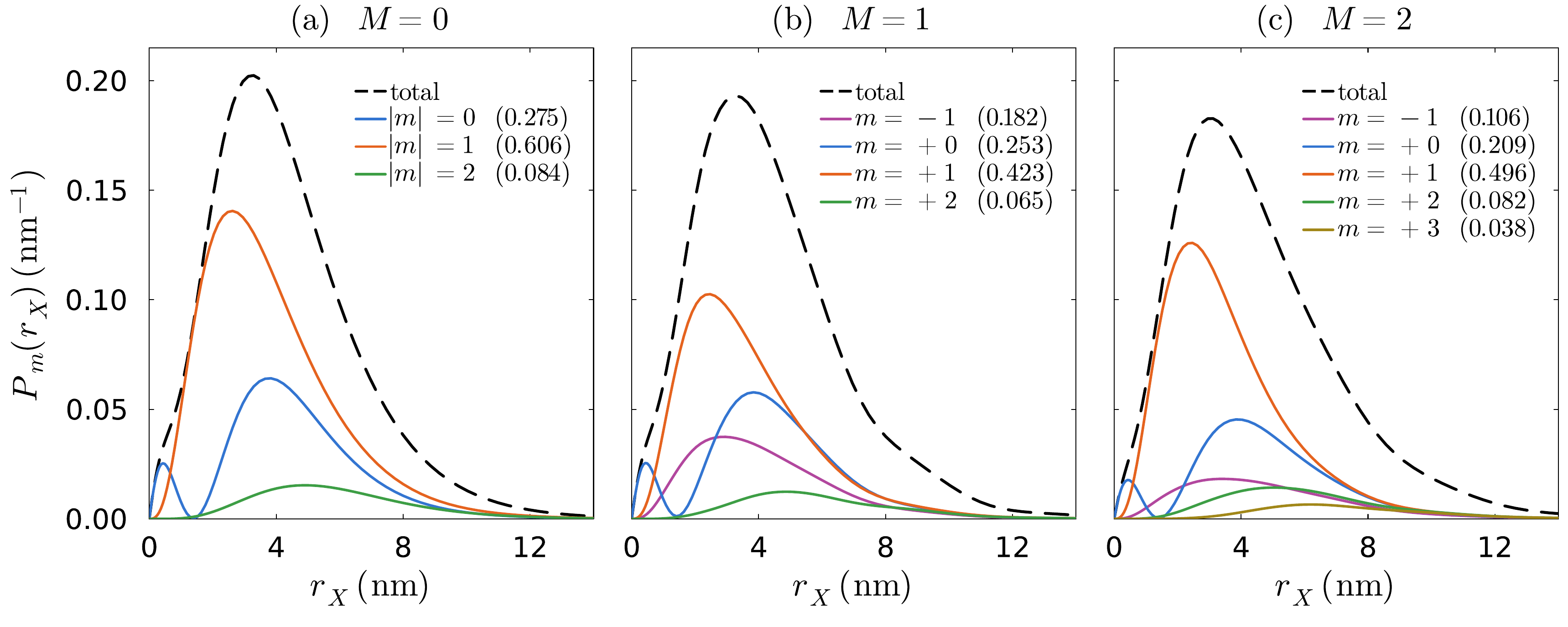}
    \caption{Radial probability density of the monolayer MoSe$_2$ Rydberg trion wave function for (a) $M=0$, (b) $M=1$ and (c) $M=2$, as a function of the bright-exciton bond length $r_X$. The black dashed line shows the total probability distribution, whereas the colored lines each indicate one angular momentum channel $m$. The number quoted in parentheses in the legend after every $m$ indicates the integrated weight of that angular momentum channel.}
    \label{fig:rad_prob_dens}
\end{figure}

The expansion in Laguerre polynomials is suitable for localized states, but not for scattering states. However, since the autodetachment linewidths of the Rydberg trion states are small, the wave functions of the $M=0-2$ trions can to a large degree be captured in this Laguerre representation: the trimer energies agree within 0.1 meV to the values obtained from the hyperspherical approach. Care needs to be taken here to choose a set of basis functions that does not lead to discretized continuum states close in energy to the quasi-bound state of interest, since this leads to level-mixing that deteriorates the description of the localized quasi-bound state wave function.

To express the wave function in terms of the bright-exciton states (where the electron and hole have the same spin and valley index), as required for computing the oscillator strength, it is most natural to switch to Jacobi coordinates: $\psi_T(r_1, r_2, \theta, \alpha) \rightarrow \psi'_T(r_X, R, \phi_X, \phi_R) $.
Here $r_X$ is the bright exciton bond length, $R$ is the exciton-electron distance, and  $\phi_X$ and $\phi_R$ represent the orientation of $r_X$ and $R$ respectively. In terms of the hole-centered coordinates (or electron-centered for positive trions) these are given by:
\begin{align}
    r_X&=r_1, \\
    R&=\sqrt{r_2^2 + \beta^2 r_1^2 - 2 \beta r_1 r_2 \cos(\theta)},&     \\
    \phi_X &= \alpha+\frac{\theta}{2} \\
    \phi_R&=\phi_X + \arctantwo \left(-r_2 \sin(\theta),r_2 \cos(\theta)-\beta r_1\right).
\end{align}
where $\beta=\frac{m_e}{m_e+m_h}$ for the negatively charged trion and $\beta=\frac{m_h}{m_e+m_h}$ for the positively charged case.  The transformation to Jacobi coordinates is performed after computing the trion wave function, since the trion calculation is more efficient in the other coordinate system.


Now we define the radial probability density
\begin{equation}
P_m(r) = r_X\int_0^{\infty} R dR\left| \int_0^{2\pi} d\phi_R \int_0^{2\pi} d\phi_X \exp(-im \phi_X) \exp(i(m-M)\phi_R) \psi_T'(r_X,R,\phi_X,\phi_R) \right|^2,
\end{equation}
which is normalized such that
\begin{equation}
    1=\sum_m \int_0^{\infty} dr_X P_m(r_X). 
\end{equation}

The angular-momentum decomposed radial probability densities for the negatively charged Rydberg trions in monolayer MoSe$_2$ are shown in Fig.~\ref{fig:rad_prob_dens}. Here we show that for each of the $M$-values, the wave function consists of an admixture of several important angular momentum components. In all these cases, the weight of the s-wave component is 20\%-28\%, while the total weight of the p-wave components lies around 60\%. 

To quantify the concrete overlap with the 2s and 2p exciton states we consider the normalized exciton wave functions $X_{nm}(r_X,\phi_X)$, which can be decomposed as
\begin{equation}
X_{nm}(r_X,\phi_X)
=
\frac{\chi_{nm}(r_X)}{\sqrt{2\pi}}
e^{im\phi_X},
\qquad
\int_0^\infty r_X\,dr_X\,\left|\chi_{nm}(r_X)\right|^2=1.
\end{equation}
For a trion with total angular momentum $M$, the amplitude in which
the bright electron--hole pair occupies the exciton state $(n,m)$ is
\begin{equation}
\mathcal{A}_{nm}(R)
=
\int_0^\infty r_X\,dr_X
\int_0^{2\pi}d\phi_X
\int_0^{2\pi}d\phi_R\,
\frac{\chi_{nm}^{*}(r_X)}{2\pi}
e^{-im\phi_X}
e^{i(m-M)\phi_R}
\psi_T'(r_X,R,\phi_X,\phi_R).
\end{equation}
The corresponding exciton-channel probability is obtained by tracing
over the radial coordinate of the spectator carrier,
\begin{equation}
P_{nm}
=
\int_0^\infty R\,dR\,
\left|\mathcal{A}_{nm}(R)\right|^2.
\end{equation}
In particular, we define
\begin{equation}
P_{2s}\equiv P_{2,0}, P_{2p}\equiv P_{2p,1}+P_{2p,-1}
\end{equation}

Concretely, for $M=0,1,2$ we respectively find \(P_{2s}=23.7\%,\ 22.0\%,\ 18.5\%\) and \(P_{2p}=50.7\%, \ 50.6\%, \ 51.6\%\). Comparing this to the weights shown in Fig.~\ref{fig:rad_prob_dens} shows that, as expected, the 2s and 2p components dominate their respective angular momentum sectors, but that other s and p states also have a small admixture. The calculations were also performed for the positively charged Rydberg trions in WSe$_2$. We find that the character of the wave functions is the same as in MoSe$_2$ up to small quantitative differences.

Finally, note that in the trion picture, which can be applied in the low-density limit, the trion oscillator strength can be calculated by computing the wave function of the trion at $r_X=0$ and the overlap of the remaining electronic wave function with the electronic wave function of the bath \cite{glazov:2020}. This overlap depends strongly on the Fermi energy and for finite $M$ this overlap is suppressed in the low-density limit. In this limit  the $M=0$ trion is therefore expected to be the brightest.

\section{Overview of samples and experimental data} \label{em:exp}

Table~\ref{tab:exp} provides an overview of all experimentally studied heterostructures and summarizes the extracted energy splittings discussed in the main text.
A detailed description of the device fabrication and characterization is given in Ref.~\cite{tuugen:2025}.

Each device consists of a WSe$_2$ layer and a MoSe$_2$ or MoS$_2$ layer separated by a thin hBN spacer with a thickness between 0.33 and 7~nm, which controls the strength of the interlayer Coulomb coupling. Note that the width of the TMD layer ($\approx$0.7 nm) needs to be added to the spacer thickness to obtain the parameter $d$ used in the main text, which specifies the distance between the centers of the TMD layers.
Top and bottom graphite gates enable independent tuning of the carrier density and the out-of-plane electric field across the heterostructure. 
The carrier density is controlled by the gate-voltage combination
\begin{equation}
V_\mu = \alpha_\mu V_\mathrm{bg} + (1 - \alpha_\mu) V_\mathrm{tg},
\end{equation}
where $V_\mathrm{tg}$ and $V_\mathrm{bg}$ denote the voltages applied to the top and bottom graphite gates, respectively. 
The coefficient $\alpha_\mu = d_\mathrm{tg} / (d_\mathrm{bg} + d_\mathrm{tg})$ accounts for the relative thicknesses of the top ($d_\mathrm{tg}$) and bottom ($d_\mathrm{bg}$) hBN dielectric layers. 

\begin{table*}[t]
    \centering
    \caption{Summary of experimental data.}
    \label{tab:exp}
    \begin{tabular*}{\textwidth}{@{\extracolsep{\fill}} c c c c c c @{}}
    \hline
    \hline
    Device & Materials & Spacer Thickness & $\Delta E_{2s,\mathrm{RAP}}$ onset & $\Delta E_{2s,\mathrm{RAP}}$ stable  & $\Delta E_{2s,1s}$ \\
    \hline
    Device 1 & WSe$_2$/MoSe$_2$ & 0.33 nm      & \(22.03 \pm 0.58\,\mathrm{meV}\)& \(20.35 \pm 0.44\,\mathrm{meV}\) & \(116.83 \pm 0.03\,\mathrm{meV}\) \\
    Device 2 & WSe$_2$/MoSe$_2$ & 0.9--1.7 nm  & \(23.88 \pm 1.51\,\mathrm{meV}\)& \(16.79 \pm 0.23\,\mathrm{meV}\) & \(120.45 \pm 0.07\,\mathrm{meV}\) \\
    Device 3 & WSe$_2$/MoS$_2$  & 1.7--2.3 nm  & \(35.66 \pm 3.31\,\mathrm{meV}\)& \(21.30 \pm 0.30\,\mathrm{meV}\)  & \(127.43 \pm 0.30\,\mathrm{meV}\) \\
    Device 4 & WSe$_2$/MoS$_2$  & 3--4 nm      & \(20.93 \pm 3.24\,\mathrm{meV}\)& \(11.21 \pm 0.22\,\mathrm{meV}\)  & \(130.48 \pm 0.13\,\mathrm{meV}\) \\
    Device 5 & WSe$_2$/MoS$_2$  & 6--7 nm      & \(10.04 \pm 3.15\,\mathrm{meV}\) & \(5.76 \pm 0.62\,\mathrm{meV}\) & \(131.48 \pm 0.45\,\mathrm{meV}\) \\
    \hline
    \hline
    \end{tabular*}
\end{table*}

Reflection spectroscopy was performed at cryogenic temperatures ($T \approx 4$~K). From the normalized reflectance spectra ($R/R_0$), we identified the neutral exciton and AP resonances in both the WSe$_2$ and MoSe$_2$/MoS$_2$ layers. The RAP feature, discussed in the main text, appears near the WSe$_2$ 2s exciton and evolves with gate voltage, as illustrated in Fig.~4a).
The peak positions of the WSe$_2$ 1s and 2s excitons and RAP were extracted by fitting the reflectance spectra at different carrier densities (see Fig.~4a), insets). From these fits, we determined the energy splitting between the 2s exciton and RAP at both the onset of the RAP and at higher densities, where the resonance energy saturates. These values are tabulated in Table~\ref{tab:exp}, along with the corresponding 2s–1s exciton splittings in WSe$_2$.

Note that the data for Device~5 are not shown in Figs.~4a) and~4b). For the corresponding hBN spacer thickness of 6–7~nm, the binding energy of the interlayer trion obtained from the theoretical calculations is negligible. While at these interlayer distances the resonance still likely follows from the hybridization of the different excitonic states due to the interaction with the electrons, the direct role of the trion in the polaron properties is reduced. For all interlayer distances, we also expect that the Berry curvature plays a role for the precise value of the RAP energy.




\end{document}